**DeepSeek reshaping healthcare in China's tertiary hospitals**

Jishizhan Chen[1], Qingzeng Zhang[2,*]

[1] Mechanical Engineering, University College London, Torrington Place, London WC1E 7JE, UK

[2] Jining Guokong Medical Co., Ltd., High-tech Zone, Guanghe Street, No. 13 Wutaizha Road, Sunshine City Garden (Area B), Jining, Shandong Province, China

[*]Correspondence: hillalien18765@gmail.com

## 1. DeepSeek deployment from Shanghai to nationwide

The integration of artificial intelligence into healthcare has accelerated in recent years, with DeepSeek emerging as a leading solution for improving clinical decision-making and hospital operational efficiency [1]. Since January 2025, the widespread adoption of this technology across China's tertiary hospitals signifies a paradigm shift in medical artificial intelligence (AI) applications. Shanghai has played a pivotal role in pioneering the implementation of DeepSeek, with leading hospitals leveraging the technology for different applications [2]. Fudan University Affiliated Huashan Hospital was one of the first to test DeepSeek 70B and its full-fledged model on multiple platforms, ensuring optimal cost-performance configurations while maintaining data security within an intranet environment. Meanwhile, Ruijin Hospital, in collaboration with Huawei, launched China's first pathology AI model, Ruizhi Pathology, which automates pathological slide analysis and has a daily processing capacity of 3,000 slides. With further multimodal integration, this system will expand to cover complex diagnostic scenarios. Similarly, Shanghai Fourth People's Hospital has implemented a localized DeepSeek deployment, integrating a medical knowledge base of over 30,000 typical cases and regional treatment guidelines, improving the efficiency of medical record generation and providing precise diagnostic support to physicians. The Jinshan Branch of Shanghai Sixth People's Hospital has fully integrated DeepSeek into physician workstations, offering real-time assistance for disease diagnosis and reducing the risk of misdiagnosis in complex cases.

Beyond Shanghai, DeepSeek has been deployed in various hospitals across China. South China Hospital in Shenzhen has developed an AI computing hub that covers clinical, research, and management functions [3]. The urology department has piloted a knowledge base assistant that allows for rapid evidence retrieval and personalized treatment recommendations. Liuzhou People's Hospital has focused on deploying DeepSeek in hematology and medical laboratories, where AI-driven cell recognition analysis has optimized the complete blood count process and improved the accuracy of microscopic blood image analysis [4]. Chengdu First People's Hospital has become the first tertiary hospital in Southwest China to complete localized DeepSeek deployment, leveraging AI for chronic disease management and telemedicine advancements [5]. Table 1 summarized China's hospitals with DeepSeek deployed for healthcare.

**Table 1.** Summary of China's hospitals with DeepSeek deployed for healthcare.

| Hospital | Location | Application | Key Benefits |
|---|---|---|---|
| Fudan University Affiliated Huashan Hospital | Shanghai | Testing DeepSeek 70B and full-fledged model | Optimizing cost-performance configurations while maintaining intranet data security |
| Ruijin Hospital | Shanghai | Pathology AI Model (Ruizhi Pathology) | Automating slide analysis, processing 3,000 slides daily, enabling multimodal diagnostic expansion |
| Shanghai Fourth People's Hospital | Shanghai | Localized medical knowledge base | Integrating 30,000+ cases, improving diagnostic precision and medical record efficiency |
| Jinshan Branch of Shanghai Sixth People's Hospital | Shanghai | AI-assisted physician workstations | Real-time assistance for disease diagnosis, reducing misdiagnosis risks |
| South China Hospital, Shenzhen | Shenzhen | AI computing hub for clinical, research, and management | Urology department AI for rapid evidence retrieval and personalized treatment |
| Liuzhou People's Hospital | Liuzhou | AI-driven hematology and medical lab systems | Automated cell recognition and blood image analysis, optimizing CBC process accuracy |
| Chengdu First People's Hospital | Chengdu | Localized AI deployment | AI-driven chronic disease management and telemedicine improvements |

## 2. Core application scenarios

DeepSeek has been integrated into several core hospital operations, particularly in intelligent pathology, imaging analysis, clinical decision-making, and workflow optimization. The implementation of AI-powered pathology and imaging analysis has led to substantial improvements in diagnostic precision and efficiency. At Ruijin Hospital, the Ruizhi Pathology AI Model automates critical pathological analysis tasks, such as tumor infiltration annotation and Ki-67 index calculation, leading to a significant increase in diagnostic efficiency [6, 7]. At Huashan Hospital, AI is being used to integrate imaging and biomarker data into a multimodal diagnostic system, where lung nodule differentiation accuracy has now surpassed 95.2%, exceeding the average performance of human doctors.

In clinical decision-making, DeepSeek is playing a crucial role in assisting physicians with real-time analysis of vast medical data. At South China Hospital in Shenzhen, DeepSeek is used to retrieve clinical evidence for urology cases, reducing the time doctors spend on literature review. Similarly, at Shanghai Fourth People's Hospital, AI-generated medical record templates allow physicians to input key patient information, with the system automatically completing 80% of the documentation. This significantly enhances administrative efficiency and allows more time for patient care.

Patient interactions and hospital workflow efficiency have also been improved with AI-driven solutions. At Jinshan Branch of Shanghai Sixth People's Hospital, an AI-powered pre-consultation system has been introduced, allowing patients to provide medical histories via AI before consulting a physician, thereby reducing outpatient waiting times. Meanwhile, Shenzhen People's Hospital has piloted an AI-powered follow-up system that utilizes sentiment analysis to provide personalized rehabilitation guidance. The system's multilingual capabilities further enhance accessibility for international patients, ensuring equitable healthcare delivery.

### 3. Technological advantages and data security

DeepSeek incorporates advanced machine learning techniques to optimize healthcare applications. The hierarchical knowledge distillation technology employed by DeepSeek enhances the balance between model generalization and precision while reducing computational inference costs by 30% [8]. Furthermore, the open-source nature of DeepSeek allows hospitals to tailor AI models according to their specific needs. For example, Liuzhou People's Hospital has developed a customized cell recognition system using DeepSeek, which is currently in its testing phase.

Ensuring data security remains a priority in AI deployment, particularly in healthcare applications where sensitive patient information is involved. All hospitals implementing DeepSeek utilize localized deployment strategies, ensuring that medical data is processed within hospital intranet environments, thereby eliminating the risks associated with external data transmission. In addition, Shanghai Fourth People's Hospital has implemented dynamic encryption and access control measures to further strengthen data security. DeepSeek deployment aligns with the Shanghai Medical AI Work Plan, which mandates that AI-generated diagnostic recommendations must be verified by physicians before being used in patient care.

### 4. Regulatory and ethical frameworks for AI-assisted diagnosis

The rapid adoption of AI in China's healthcare sector is driven by strong policy momentum [9]. However, the increasing reliance on AI-driven diagnostics raises critical ethical and regulatory concerns. The rapid deployment of DeepSeek across Chinese hospitals, often ahead of comprehensive national regulations, has reignited debates on accountability in AI-assisted diagnosis. While AI models currently handle fundamental tasks like medical knowledge retrieval, their integration with patient data—including imaging and lab results—raises pressing questions: who is responsible for AI errors, and how should conflicts between human clinicians and AI predictions be adjudicated?

This issue stems from a systemic 'responsibility gap' where technological innovation outpaces governance. Developers often argue that AI is merely a decision-support tool, clinicians emphasize that ultimate accountability lies with human oversight, and regulators struggle to define robust liability frameworks. For instance, in the UK, AI-driven stroke diagnosis has transformed workflows by prioritizing algorithmic predictions [10], yet clinicians maintain oversight through iterative verification. However, such hybrid models also introduce risks of automation bias, where overburdened clinicians may defer excessively to AI-generated recommendations. To bridge this gap, an integrated regulatory approach is necessary. Policymakers must establish dynamic accountability frameworks that define shared responsibility: developers should ensure algorithmic transparency and robustness, clinicians must retain override authority, and healthcare institutions need to implement audit protocols to ensure responsible AI deployment. Additionally, AI systems should incorporate real-time explainability mandates, requiring them to disclose context-specific rationales, such as data limitations, to assist clinicians in making informed decisions.

Furthermore, liability structures must evolve to accommodate AI's growing role in diagnostics. A promising approach is the creation of liability insurance pools jointly funded by AI vendors and hospitals, providing compensation for AI-related errors without discouraging innovation. This model, successfully tested in Germany's autonomous vehicle sector, could offer a viable solution for AI-assisted diagnosis. Without these measures, unregulated AI applications in healthcare risk eroding public trust and impeding broader adoption.

Industry collaborations also play a pivotal role in shaping the future of AI in healthcare. The National Health Commission has mandated that by 2025, all tertiary hospitals must integrate AI-assisted diagnosis [11], prompting institutions such as Chengdu First People's Hospital to embed AI into their performance evaluation frameworks. Additionally, hospitals are partnering with technology leaders like Huawei and SenseTime to develop robust AI infrastructure. Academic institutions, such as Shanghai Jiao Tong University, have also deployed DeepSeek on high-performance computing platforms to enhance AI-driven medical education and research [12]. Moreover, Ruijin Hospital has spearheaded an Innovation Laboratory that fosters cross-hospital research collaborations, inviting institutions nationwide to participate in AI-driven medical advancements [13].

A well-regulated AI ecosystem will be essential for ensuring AI's long-term success in medical diagnostics. By establishing clear accountability mechanisms, improving transparency, and fostering industry collaboration, healthcare systems can harness AI's potential while mitigating ethical and legal risks. The challenge now lies in developing policies that strike a balance between innovation and responsibility, ensuring that AI serves as a reliable partner rather than an unchecked decision-maker in clinical settings.

**5. Future prospects: towards comprehensive AI integration**

The future of AI in healthcare will see greater integration of multimodal data sources. Ruijin Hospital is currently developing an AI model that fuses genomics and radiomics data, paving the way for precision medicine and personalized treatments. Additionally, Huashan Hospital is piloting a human-machine hybrid outpatient system, which is expected to improve waiting efficiency by 300% by streamlining patient management processes.

Beyond major medical centers, AI applications are set to expand into smaller regional hospitals. Liuzhou People's Hospital, for example, aims to extend its AI applications to county-level hospital networks, ensuring that high-quality AI-driven healthcare services reach broader populations. As AI continues to evolve, it will play an increasingly central role in bridging the gap between urban and rural healthcare facilities, improving accessibility, and enhancing patient outcomes nationwide.

The large-scale deployment of DeepSeek in China's tertiary hospitals signifies a transition from technological validation to real-world implementation. By driving clinical assistance, process optimization, and data security frameworks, hospitals are reshaping medical service efficiency and quality. With continued policy support, industry collaboration, and technological advancements, AI is poised to become a core driver of medical innovation, fundamentally transforming the landscape of healthcare delivery in China and beyond.

**Authors' Contributions:**

J.C. & Q.Z.: conceptualisation, writing – original draft, writing – review & editing.

**Conflict of Interest Statements:**

We declare no competing interests.

**Role of Funding Source:**

No funding was received for this study.

**Ethics Committee approval:**
N/A.